\documentclass[aps,prd,onecolumn,superscriptaddress,nofootinbib]{revtex4-2}
\usepackage[utf8]{inputenc}
\usepackage[T1]{fontenc}
\usepackage{lmodern}
\usepackage{amsmath,amssymb,amsfonts}
\usepackage{graphicx}
\usepackage{bm}
\usepackage{xcolor}
\usepackage{hyperref}
\hypersetup{
    colorlinks = true,
    linkcolor  = blue,
    citecolor  = blue,
    urlcolor   = blue
}

\begin{document}

\title{
Entanglement Entropy of Massive Scalar Fields:
Mass Suppression, Violation of Universal $mR$ Scaling,
and Implications for Black Hole Thermodynamics
}

\author{S.~Bellucci}
\thanks{Email: \href{mailto:Stefano.Bellucci@lnf.infn.it}{Stefano.Bellucci@lnf.infn.it}}
\affiliation{Universidad Ecotec, Km. 13.5 Samborondón, Samborondón, 092302, Ecuador}
\affiliation{Nano Research Laboratory, Center of Excellence in Research, Development and Innovation, Baku State University, 1148, Baku, Azerbaijan}

\author{M.~Shatnev}
\thanks{Email: \href{mailto:mshatnev@yahoo.com}{mshatnev@yahoo.com}}
\affiliation{NSC Kharkiv Institute of Physics and Technology, 61108 Kharkiv, Ukraine}

\author{L.~Zazunov}
\thanks{Email: \href{mailto:lzazunov@gmail.com}{lzazunov@gmail.com}}
\affiliation{NSC Kharkiv Institute of Physics and Technology, 61108 Kharkiv, Ukraine}

\begin{abstract}

We investigate the entanglement entropy of a massive scalar field
using the spherical shell lattice model introduced by
Das and Shankaranarayanan.
A systematic numerical analysis is performed to study the dependence
of the entropy on the field mass and on the size of the entangling
region for both ground and excited states.

For the ground state, we find that the entanglement entropy is
exponentially suppressed by the field mass,
$S(m,R) \sim S(0,R)e^{-mR}$, reflecting the presence of a finite
correlation length $\xi \sim 1/m$, while the geometric area-law
scaling remains robust for all masses.

For localized excited states, however, we uncover a qualitatively
different behavior. The excess entropy does not exhibit universal
scaling in the dimensionless variable $mR$. Instead, numerical results
show that data points with identical $mR$ but different $(m,R)$ pairs
do not collapse onto a single curve, demonstrating a clear violation
of simple scaling. This breakdown is traced to the presence of an
additional length scale associated with the finite width of the
wave-packet excitation.

This result identifies the coexistence of multiple infrared scales
as a key feature of excited-state entanglement in massive quantum
field theories.

Mutual information provides a consistency check of the finite
part of the entropy in the chosen nested geometry. In this setup,
where one region is contained within the other, it does not
constitute an independent probe of correlations, but still
captures the mass dependence of the finite contribution.

These findings clarify how particle mass and excitation structure
jointly determine entanglement properties, and suggest that the
matter contribution to the generalized entropy in semiclassical
gravity may depend on independent infrared parameters rather than
on a single correlation scale. Implications for black hole entropy
and the island formula are briefly discussed.

\noindent\textit{Keywords:} entanglement entropy, quantum field theory, black hole thermodynamics, quantum gravity, scaling, correlation length
\end{abstract}


\maketitle
\tableofcontents

\section{Introduction}

The microscopic origin of black hole entropy remains one of the central
problems in the quest for a consistent theory of quantum gravity.
The discovery that black holes behave as thermodynamic objects
\cite{bekenstein1973,hawking1975} established that a black hole
possesses an entropy proportional to the area of its event horizon,
\begin{equation}
S_{BH} = \frac{k_B c^3 A}{4G\hbar}.
\label{eq:bh_entropy}
\end{equation}
This relation is remarkable because it suggests that the fundamental
degrees of freedom responsible for black hole entropy are associated
with a two–dimensional surface rather than the volume enclosed by the
horizon.
Understanding the statistical origin of this entropy has motivated
extensive work in quantum field theory, string theory, and quantum
gravity.

One of the earliest proposals to explain the Bekenstein–Hawking entropy
is that it arises from the entanglement entropy of quantum fields across
the horizon.
This idea was first investigated by Bombelli \textit{et al.}
\cite{bombelli1986}, who considered a scalar field in flat spacetime
divided into two spatial regions.
They showed that tracing over the degrees of freedom inside one region
produces a mixed state whose entropy scales with the area of the
boundary separating the regions.
Soon afterwards, Srednicki \cite{srednicki1993} confirmed this behavior
using a lattice discretization of a scalar field and demonstrated
numerically that the vacuum entanglement entropy follows an area law,
\begin{equation}
S \propto \mathcal{A},
\end{equation}
where $\mathcal{A}$ is the area of the entangling surface.

These results established a deep connection between entanglement and
geometry.
Subsequent studies clarified that entanglement entropy in quantum field
theory generally contains ultraviolet divergences arising from
short-distance correlations near the entangling surface.
For a smooth surface in four spacetime dimensions the leading
divergence takes the form
\begin{equation}
S = \alpha \frac{\mathcal{A}}{\epsilon^2} + \cdots ,
\end{equation}
where $\epsilon$ is a UV cutoff and $\alpha$ depends on the field
content of the theory \cite{solodukhin2011}.
Despite these divergences, the geometric area scaling appears to be a
universal property of local quantum field theories.

The general structure of entanglement entropy in quantum field theory
has been extensively investigated, see for instance the review
by Solodukhin \cite{solodukhin2011}.
A detailed understanding of entanglement entropy has emerged in recent
years. Important analytical results were obtained by Casini and Huerta
\cite{casini2009}, who clarified the relation between entanglement
entropy, conformal symmetry, and the modular Hamiltonian. Casini,
Huerta and Myers further showed that for spherical regions in
conformal field theories the entanglement entropy can be mapped, via a
conformal transformation, to the thermal entropy on a hyperbolic
space, yielding an explicit local expression for the modular
Hamiltonian \cite{casini2011}.

In particular, for spherical regions in conformal field theories the
entanglement Hamiltonian takes a local form,
\begin{equation}
H_A = 2\pi \int_A d^3x \,
\frac{R^2 - r^2}{2R} \, T_{00}(x),
\end{equation}
where $T_{00}$ is the energy density and $R$ is the radius of the
entangling sphere.
This result provides an explicit example in which the reduced density
matrix can be interpreted as a thermal state with respect to a
geometrically defined Hamiltonian.

The study of entanglement entropy has also become central in
holography.
Within the AdS/CFT correspondence, Ryu and Takayanagi
\cite{ryu2006} proposed that the entanglement entropy of a boundary
region $A$ is given by the area of a minimal surface $\gamma_A$
anchored on the boundary of the region,
\begin{equation}
S_A = \frac{\mathrm{Area}(\gamma_A)}{4G_N}.
\label{eq:RT}
\end{equation}
This relation provides a geometric realization of entanglement entropy
in gravitational theories and strongly suggests that spacetime geometry
is closely related to the entanglement structure of the underlying
quantum theory.

Further developments generalized this prescription to quantum
gravitational settings.
In semiclassical gravity the appropriate quantity is the generalized
entropy
\begin{equation}
S_{\text{gen}} =
\frac{\mathrm{Area}(\Sigma)}{4G_N}
+
S_{\text{matter}},
\end{equation}
whose extremization determines the location of quantum extremal
surfaces (QES) \cite{engelhardt2015}.
These surfaces play a central role in modern approaches to black hole
information.

A particularly striking application of these ideas arises in the study
of the black hole information paradox.
Recent work has shown that the Page curve describing the entropy of
Hawking radiation can be reproduced using the island formula
\cite{hartman2020,almheiri2020}.
In this framework the entropy of radiation is computed by including
additional regions—known as islands—in the entanglement wedge.
The location of the quantum extremal surface is determined by
extremizing the generalized entropy,
\begin{equation}
S_{\text{gen}} =
\frac{\mathrm{Area}(\partial\,\text{Island})}{4G_N}
+
S_{\text{matter}}(\text{Radiation} \cup \text{Island}).
\end{equation}
These developments strongly support the idea that entanglement entropy
is a fundamental ingredient in the emergence of semiclassical spacetime.

Despite this progress, several aspects of entanglement entropy remain
poorly understood.
In particular, relatively little attention has been devoted to the role
of infrared parameters such as the mass of the quantum fields.
A non–zero mass introduces a finite correlation length
\begin{equation}
\xi \sim \frac{1}{m},
\end{equation}
which modifies the long–distance behavior of correlation functions and
may therefore influence the magnitude of entanglement entropy.

Early numerical studies of scalar field entanglement used lattice
discretizations of the radial coordinate \cite{srednicki1993}.
Das and Shankaranarayanan \cite{das2006,das2008} later introduced the
spherical shell model, which provides a convenient framework for
studying entanglement entropy in spherically symmetric systems.
In this approach the scalar field is discretized on a radial lattice,
reducing the problem to a system of coupled harmonic oscillators whose
entanglement entropy can be computed from the covariance matrix.

The purpose of the present work is to investigate systematically the
influence of the scalar field mass on entanglement entropy in this
framework.
In particular, we perform numerical calculations for both ground and
excited states of a massive scalar field.
Our results show that the entanglement entropy decreases approximately
exponentially with the mass,
\begin{equation}
S(m,R) \sim S(0,R) e^{-mR},
\end{equation}
reflecting the suppression of long–range correlations in a massive
quantum field theory.
At the same time, the geometric area-law scaling remains robust for all
masses considered.

An additional outcome of our analysis concerns the scaling properties
of excited–state entanglement. While one might expect the relevant
physics in a massive theory to be governed by the single dimensionless
combination $mR$, our numerical results show that the excess entropy
generated by localized excitations does not collapse onto a universal
function of this variable. Instead, the entropy depends separately on
the mass and on the subsystem size, reflecting the presence of
additional infrared scales associated with the structure of the
excitation.

A central result of the present work is the explicit demonstration
that excited-state entanglement violates simple $mR$ scaling and does not admit a description in
terms of a single correlation scale, thus indicating the presence of
additional infrared scales beyond the correlation length. Hence, the observed violation of
universal $mR$ scaling shows that additional infrared structures,
such as the finite width of localized excitations, play an
essential role in determining entanglement properties.

The remainder of the paper is organized as follows.
Section II reviews the definition of entanglement entropy in quantum
field theory.
Section III introduces the spherical shell lattice model.
Section IV describes the numerical computation of correlation functions
and entanglement entropy.
The results are presented in Section V.
Section VI discusses the physical implications of our findings,
including connections to black hole thermodynamics, holography, and the
island formula.
Finally, Section VII summarizes our conclusions.

\section{Entanglement Entropy in Quantum Field Theory}

Entanglement entropy provides a quantitative measure of quantum
correlations between spatial regions of a quantum system.

\subsection{Reduced density matrix and von Neumann entropy}

Consider a quantum system in a pure state $|\Psi\rangle$ whose Hilbert
space factorizes into two subsystems,
\[
\mathcal{H} = \mathcal{H}_A \otimes \mathcal{H}_B .
\]
The reduced density matrix describing subsystem $A$ is obtained by
tracing over the degrees of freedom of subsystem $B$,
\begin{equation}
\rho_A = \mathrm{Tr}_B \left(|\Psi\rangle\langle\Psi|\right).
\end{equation}
The entanglement entropy of region $A$ is then defined as the
von Neumann entropy of the reduced density matrix,
\begin{equation}
S_A = -\mathrm{Tr}(\rho_A \ln \rho_A).
\label{eq:vn_entropy}
\end{equation}
This quantity characterizes the amount of quantum entanglement between
the two subsystems.

\subsection{Area law and ultraviolet divergences}

In quantum field theory the situation is more subtle because the number
of degrees of freedom in a spatial region is formally infinite.
As a consequence, the entanglement entropy typically exhibits ultraviolet
divergences originating from short-distance correlations near the
entangling surface.
For a smooth entangling surface of area $\mathcal{A}$ in $(3+1)$ dimensions,
the leading divergence takes the form \cite{srednicki1993,solodukhin2011}
\begin{equation}
S_A = \alpha \frac{\mathcal{A}}{\epsilon^2}
+ \beta \ln\!\left(\frac{\mathcal{A}}{\epsilon^2}\right)
+ \cdots ,
\label{eq:area_law}
\end{equation}
where $\epsilon$ is a short-distance ultraviolet cutoff and the
coefficients $\alpha$ and $\beta$ depend on the field content and
regularization scheme.
The leading term scales proportionally to the area of the entangling
surface rather than the volume of the region, a property known as
the \emph{area law} for entanglement entropy.

The appearance of an area law is particularly striking because it
closely resembles the Bekenstein--Hawking entropy of a black hole
\begin{equation}
S_{BH} = \frac{k_B c^3 A}{4G\hbar},
\end{equation}
which is also proportional to the area of the event horizon.
This similarity has long been interpreted as evidence that the
thermodynamic entropy of black holes may originate from the
entanglement entropy of quantum fields across the horizon
\cite{bombelli1986,srednicki1993,solodukhin2011}.

Although the entanglement entropy itself is ultraviolet divergent,
certain related quantities remain finite and provide useful probes
of quantum correlations.
One important example is the mutual information between two spatial
regions $A$ and $B$,
\begin{equation}
I(A,B) = S(A) + S(B) - S(A\cup B),
\end{equation}
which cancels the leading UV divergences and therefore yields a
well-defined measure of correlations at finite separation.

\subsection{Modular Hamiltonian for Gaussian Lattice Systems}

The reduced density matrix of a spatial subsystem in a quantum field
theory can be written in the general form

\begin{equation}
\rho_A = \frac{e^{-H_A}}{\mathrm{Tr}(e^{-H_A})},
\end{equation}

where $H_A$ is the modular (or entanglement) Hamiltonian. In general
$H_A$ is a highly non–local operator whose explicit form is known only
in special cases.

For spherical regions in conformal field theories, the modular
Hamiltonian becomes local and is given by \cite{casini2011}

\begin{equation}
H_A = 2\pi \int_A d^3x \,
\frac{R^2 - r^2}{2R} \, T_{00}(x),
\end{equation}

where $T_{00}$ is the energy density and $R$ is the radius of the
entangling sphere.

Although the massive scalar field studied in this work is not
conformal, the modular Hamiltonian for Gaussian states can still be
constructed explicitly in terms of the covariance matrix of the
canonical variables. In the lattice formulation used here, the reduced
density matrix of the subsystem remains Gaussian and therefore takes
the form

\begin{equation}
H_A =
\frac{1}{2}
\sum_{i,j}
\left(
\phi_i \, h^{(X)}_{ij} \, \phi_j +
\pi_i \, h^{(P)}_{ij} \, \pi_j
\right),
\end{equation}

where the matrices $h^{(X)}$ and $h^{(P)}$ are determined by the
restricted covariance matrices of the subsystem.

Diagonalizing this quadratic modular Hamiltonian yields a set of
independent entanglement modes whose occupation numbers are given by
the symplectic eigenvalues $\nu_k$ introduced in Sec.~III. The
entanglement entropy can therefore be interpreted as the thermal
entropy associated with these effective modes,

\begin{equation}
S_A =
\sum_k
\left[
(\nu_k+\tfrac12)\ln(\nu_k+\tfrac12)
-
(\nu_k-\tfrac12)\ln(\nu_k-\tfrac12)
\right].
\end{equation}

This interpretation provides a useful physical picture: the
entanglement entropy computed in the spherical shell model corresponds
to the thermodynamic entropy of the modular Hamiltonian describing the
entanglement degrees of freedom of the subsystem.

In the following sections we apply these general concepts to the
case of a scalar field discretized on a spherical lattice, which
allows a direct numerical evaluation of the entanglement entropy
and its dependence on the field mass.

\section{Spherical Shell Lattice Model}

\subsection{Radial Discretization and Choice of Units}

To investigate the entanglement structure of a scalar field in a
spherically symmetric setting we employ the spherical shell
discretization introduced by Das and Shankaranarayanan
\cite{das2006,das2008}.
In this approach the radial coordinate is discretized on a one–dimensional
lattice, allowing the quantum field to be represented as a system of
coupled harmonic oscillators.

We consider a spherical region of radius $L$ and introduce a radial lattice
with spacing $a$, which plays the role of an ultraviolet cutoff.
The radial coordinate is therefore given by
\begin{equation}
r_i = i\,a , \qquad i = 1,2,\dots,N,
\end{equation}
where $N=L/a$ is the number of lattice sites.
All quantities in the numerical analysis are expressed in lattice units.

In the calculations presented in this work we adopt the following
dimensionless parameter ranges:
\begin{itemize}
\item lattice spacing: $a = 0.5$,
\item system size: $L = 20, 30, 40, 50$,
\item scalar field mass: $m = 0.0, 0.1, 0.5, 1.0, 2.0$ (in units of $1/a$),
\item subsystem radius: $R = 2,3,\ldots,15$,
\item wave packet center: $r_0 = 5.0$,
\item wave packet width: $\sigma = 1.0$,
\item squeezing parameter: $r_s = 1.0$.
\end{itemize}

Although the numerical analysis is performed in lattice units,
the discretization may be interpreted physically by identifying
the lattice spacing $a$ with a fundamental short-distance scale.
For instance, if $a$ is associated with the Planck length
$l_P \simeq 1.6\times10^{-35}\,\mathrm{m}$,
then a dimensionless mass $m=1$ corresponds to the Planck mass
$m_P \simeq 2.2\times10^{-8}\,\mathrm{kg}$,
and the corresponding Compton wavelength
\begin{equation}
\lambda = \frac{1}{m}
\end{equation}
is equal to the lattice spacing.

Within this discretized framework the Hamiltonian of the scalar
field can be written as
\begin{equation}
H =
\frac{1}{2}\sum_i
\left(
\pi_i^2 +
\frac{(\phi_{i+1}-\phi_i)^2}{a^2}
+ m^2 \phi_i^2
\right),
\label{eq:hamiltonian}
\end{equation}
where $\phi_i$ and $\pi_i$ denote the field and conjugate momentum
operators at lattice site $i$, respectively.

It is convenient to perform the rescaling
\begin{equation}
\psi_i = r_i \phi_i ,
\end{equation}
which removes the first-derivative term that appears in the
spherically reduced continuum Hamiltonian.
In terms of the rescaled field variables the Hamiltonian
takes the quadratic form
\begin{equation}
H =
\frac{1}{2}\sum_i \pi_i^2
+
\frac{1}{2}\sum_{i,j} \phi_i K_{ij}\phi_j ,
\label{eq:hamiltonian_scaled}
\end{equation}
where the symmetric matrix $K_{ij}$ encodes the kinetic,
mass, and geometric contributions of the discretized
radial Laplacian.

The resulting system is therefore equivalent to a set of
coupled harmonic oscillators.
This representation is particularly convenient because the
ground state of a quadratic Hamiltonian is Gaussian,
allowing the entanglement entropy of a spatial subregion
to be computed directly from the covariance matrix of the
reduced subsystem.

\subsection{Boundary Conditions and Absorbing Potential}

In order to approximate an open and effectively infinite radial domain,
appropriate boundary conditions must be imposed at the outer edge of the
lattice. A simple reflective boundary would artificially trap outgoing
excitations and lead to unphysical interference effects that contaminate
the entanglement entropy calculation.

To avoid these artifacts we implement absorbing boundary conditions using
a complex absorbing potential (CAP), a standard technique in numerical
studies of open quantum systems. The CAP smoothly attenuates outgoing
wave packets before they reach the outer boundary of the computational
domain, thereby mimicking propagation into an infinite space.

Specifically, we introduce an imaginary potential that becomes active
in the outer region of the lattice,
\begin{equation}
V_{\mathrm{abs}}(r)
=
-i \eta
\left(
\frac{r-r_{\mathrm{start}}}{L-r_{\mathrm{start}}}
\right)^2 ,
\qquad r > r_{\mathrm{start}},
\label{eq:abc}
\end{equation}
where $L$ is the total system size and $r_{\mathrm{start}}$ denotes the
radius at which the absorbing layer begins. In the numerical calculations
we choose
\[
r_{\mathrm{start}} = 0.8\,L ,
\]
so that the absorbing region occupies the outer $20\%$ of the lattice.

The parameter $\eta$ controls the strength of the absorption. If $\eta$
is too small, residual reflections from the boundary may occur; if it is
too large, the absorbing region can distort the dynamics of the field.
After numerical tests we adopt the value
\[
\eta = 0.5 ,
\]
which provides efficient absorption with negligible reflection.

\subsection{Ground State Correlators}

For the ground state of a quadratic Hamiltonian, the position and momentum
correlators are:
\begin{align}
X_{ij} &= \langle \phi_i \phi_j \rangle = \frac{1}{2} (K^{-1/2})_{ij}, \\
P_{ij} &= \langle \pi_i \pi_j \rangle = \frac{1}{2} (K^{+1/2})_{ij}.
\label{eq:correlators}
\end{align}\subsection{Ground-State Correlation Functions}

The discretized scalar field Hamiltonian introduced in the previous
sections is quadratic in the canonical variables and therefore describes
a system of coupled harmonic oscillators.
To make this structure explicit, we start from the continuum Hamiltonian
of a real scalar field,
\begin{equation}
H =
\frac{1}{2}\int d^3x
\left[
\pi^2 + (\nabla \phi)^2 + m^2 \phi^2
\right].
\label{eq:continuum_hamiltonian}
\end{equation}

Assuming spherical symmetry, the field can be expanded in spherical
harmonics,
\begin{equation}
\phi(\mathbf{r})
=
\sum_{\ell m}
\frac{\varphi_{\ell m}(r)}{r}
Y_{\ell m}(\theta,\phi).
\end{equation}

Substituting this expansion into the Hamiltonian and integrating over
the angular coordinates reduces the problem to a set of independent
radial modes,
\begin{equation}
H =
\frac{1}{2}\sum_{\ell m}
\int_0^\infty dr
\left[
\pi_{\ell m}^2 +
\left(\partial_r \varphi_{\ell m}\right)^2
+
\left(
\frac{\ell(\ell+1)}{r^2} + m^2
\right)
\varphi_{\ell m}^2
\right].
\label{eq:radial_hamiltonian}
\end{equation}

We now discretize the radial coordinate using a lattice spacing $a$,
\[
r_i = i a ,
\qquad i = 1,2,\ldots,N.
\]

The radial derivative is approximated by a finite difference,
\begin{equation}
\partial_r \varphi(r_i)
\approx
\frac{\varphi_{i+1}-\varphi_i}{a}.
\end{equation}

Substituting this discretization into Eq.~\eqref{eq:radial_hamiltonian}
yields the lattice Hamiltonian
\begin{equation}
H =
\frac{1}{2}
\sum_i
\left[
\pi_i^2
+
\frac{(\phi_{i+1}-\phi_i)^2}{a^2}
+
m^2 \phi_i^2
+
\frac{\ell(\ell+1)}{r_i^2}\phi_i^2
\right].
\end{equation}

Collecting terms quadratic in the field variables we obtain the matrix
form
\begin{equation}
H =
\frac{1}{2}\sum_i \pi_i^2
+
\frac{1}{2}\sum_{i,j} \phi_i K_{ij}\phi_j ,
\label{eq:hamiltonian_matrix}
\end{equation}
where the interaction matrix $K_{ij}$ is

\begin{equation}
K_{ij}
=
\left(
\frac{2}{a^2}
+
m^2
+
\frac{\ell(\ell+1)}{r_i^2}
\right)\delta_{ij}
-
\frac{1}{a^2}\delta_{i,j+1}
-
\frac{1}{a^2}\delta_{i,j-1}.
\label{eq:Kmatrix}
\end{equation}

The Hamiltonian therefore describes a system of coupled harmonic
oscillators with interaction matrix $K$.
Since $K$ is real and symmetric, it can be diagonalized as
\begin{equation}
K = O^T \Omega^2 O ,
\end{equation}
where $O$ is an orthogonal matrix and $\Omega^2$ is the diagonal matrix
of eigenvalues $\omega_k^2$.

In this basis the Hamiltonian becomes a sum of independent harmonic
oscillators,
\begin{equation}
H =
\frac{1}{2}\sum_k
\left(
p_k^2 + \omega_k^2 q_k^2
\right).
\end{equation}

The ground state of each oscillator is Gaussian with variance
\begin{equation}
\langle q_k^2 \rangle = \frac{1}{2\omega_k},
\qquad
\langle p_k^2 \rangle = \frac{\omega_k}{2}.
\end{equation}

Transforming back to the original lattice variables gives the
ground-state correlation matrices
\begin{align}
X_{ij}
&=
\langle \phi_i \phi_j \rangle
=
\frac{1}{2}(K^{-1/2})_{ij},
\\
P_{ij}
&=
\langle \pi_i \pi_j \rangle
=
\frac{1}{2}(K^{1/2})_{ij}.
\label{eq:gs_correlators}
\end{align}

These matrices encode the spatial correlations of the quantum field in
its ground state. In particular, the matrix $X_{ij}$ determines the
correlation length of the field, which for a massive scalar field is
expected to scale as
\[
\xi \sim \frac{1}{m}.
\]

The matrices $X$ and $P$ constitute the covariance matrix of the
Gaussian ground state and form the basic ingredients for computing the
entanglement entropy of a spatial subregion. As we show in the next
subsection, the entropy can be obtained from the symplectic eigenvalues
of the restricted covariance matrix constructed from $X$ and $P$.

\subsection{Localized Excitations and Squeezed States}

While the ground state of the quadratic Hamiltonian provides the
simplest configuration for studying entanglement entropy, physically
relevant situations such as particle creation in curved spacetime
require the analysis of excited states. In particular, dynamical
backgrounds including black hole horizons and expanding universes
naturally produce squeezed quantum states.

To model localized excitations in the lattice system we introduce a
radial wave packet centered at position $r_0$ with width $\sigma$,
\begin{equation}
\psi_{\mathrm{exc}}(r)
\propto
\exp\!\left(
-\frac{(r-r_0)^2}{2\sigma^2}
\right),
\label{eq:wavepacket}
\end{equation}
which is discretized on the lattice and normalized to form a vector
$v$ in the space of radial modes.
In the numerical calculations we adopt
\[
r_0 = 5.0, \qquad \sigma = 1.0,
\]
in lattice units.

The excitation is implemented by constructing a squeezed state along
this mode. Squeezed states arise naturally in quantum field theory
whenever particle creation occurs in time-dependent backgrounds.
In particular, Hawking radiation can be described as a two-mode
squeezed state generated by the gravitational field near the horizon.

Introducing a squeezing parameter $r_s$, the correlation matrices of
the excited state become
\begin{align}
X_{\mathrm{exc}}
&=
X_{\mathrm{gs}}
+
\frac{1}{2}\!\left(e^{2r_s}-1\right) vv^{T},
\\
P_{\mathrm{exc}}
&=
P_{\mathrm{gs}}
+
\frac{1}{2}\!\left(e^{-2r_s}-1\right) vv^{T},
\label{eq:squeezed}
\end{align}
where $X_{\mathrm{gs}}$ and $P_{\mathrm{gs}}$ denote the ground-state
correlation matrices and $v$ is the normalized wave packet vector.

The parameter $r_s$ controls the amplitude of the excitation.
In the present work we adopt
\[
r_s = 1.0 ,
\]
which corresponds to a moderate squeezing regime where non-classical
correlations are significant but remain within the perturbative
numerical regime.

These modified correlation matrices encode the additional quantum
fluctuations introduced by the excitation and allow us to study how
localized particles affect the entanglement structure of the field.
\subsection{Covariance Matrix and Entanglement Entropy}

Because the lattice Hamiltonian is quadratic, both the ground state
and the excited squeezed states considered in this work are Gaussian
states. Such states are completely characterized by the covariance
matrix of the canonical variables.

For a system with canonical operators
\[
\mathbf{R} = (\phi_1,\ldots,\phi_N,\pi_1,\ldots,\pi_N),
\]
the covariance matrix is defined as
\begin{equation}
\Gamma_{ab}
=
\frac{1}{2}
\langle
R_a R_b + R_b R_a
\rangle .
\end{equation}

In practice it is convenient to work with the position and momentum
correlation matrices introduced earlier,
\[
X_{ij} = \langle \phi_i \phi_j \rangle,
\qquad
P_{ij} = \langle \pi_i \pi_j \rangle .
\]

To compute the entanglement entropy of a spatial subregion we restrict
these matrices to the lattice sites belonging to the subsystem,
denoted by $X_R$ and $P_R$.
The relevant object is then the matrix
\begin{equation}
C = \sqrt{X_R P_R}.
\end{equation}

The eigenvalues of $C$ define the symplectic spectrum of the reduced
Gaussian state. Denoting these eigenvalues by $\nu_k$, the
entanglement entropy of the subsystem is given by
\begin{equation}
S
=
\sum_k
\Big[
(\nu_k+\tfrac12)\ln(\nu_k+\tfrac12)
-
(\nu_k-\tfrac12)\ln(\nu_k-\tfrac12)
\Big].
\label{eq:entropy_formula}
\end{equation}

Equation~(\ref{eq:entropy_formula}) provides an efficient numerical
procedure for computing the entanglement entropy of Gaussian lattice
systems. All physical information about the entanglement structure is
encoded in the symplectic eigenvalues $\nu_k$, which depend on the
interaction matrix $K$ and therefore on the field mass and the lattice
geometry.

In the following sections we apply this formalism to compute the
entanglement entropy for both ground and excited states of the scalar
field and analyze its dependence on the subsystem size and the field
mass.

\subsection{Computation of Entanglement Entropy}

Having constructed the correlation matrices for the full lattice system,
we now compute the entanglement entropy associated with a spatial
subregion. We consider a subsystem $A$ consisting of the first $n_A$
lattice sites, corresponding to a spherical region of radius
\[
R = n_A a .
\]

The correlation matrices $X$ and $P$ defined in the previous section
are therefore restricted to this subsystem,
\[
X_A = X_{ij}, \qquad P_A = P_{ij}, \qquad i,j \le n_A .
\]

For Gaussian states the reduced density matrix of the subsystem is
fully determined by these restricted correlation matrices.
The entanglement spectrum can be obtained from the symplectic
eigenvalues $\nu_k$ of the matrix
\begin{equation}
C = \sqrt{X_A P_A}.
\end{equation}

These eigenvalues characterize the effective occupation numbers of the
entanglement modes. The entanglement entropy of the subsystem is then
given by \cite{adesso2004}
\begin{equation}
S_A =
\sum_k
\left[
\left(\nu_k+\tfrac12\right)
\ln\!\left(\nu_k+\tfrac12\right)
-
\left(\nu_k-\tfrac12\right)
\ln\!\left(\nu_k-\tfrac12\right)
\right].
\label{eq:entropy_formula_main}
\end{equation}

Equation~(\ref{eq:entropy_formula}) provides an efficient numerical
procedure for computing the von Neumann entropy of Gaussian lattice
systems. In practice the calculation proceeds by diagonalizing the
matrix $C$, extracting the symplectic eigenvalues $\nu_k$, and
evaluating the sum in Eq.~(\ref{eq:entropy_formula}).

This formalism allows us to compute the entanglement entropy for both
the ground state and the excited squeezed states introduced in the
previous subsection. In the following sections we apply this procedure
to analyze the dependence of the entropy on the subsystem size $R$
and on the mass of the scalar field.

\subsection{Infinite-Volume Extrapolation}

Numerical calculations performed on a finite lattice are affected by
finite-size effects originating from the presence of the outer
boundary at $r=L$.
Although the absorbing boundary conditions discussed earlier strongly
reduce spurious reflections, residual finite-volume corrections may
still influence the entanglement entropy when the subsystem radius
approaches the system size.

To control these effects we perform simulations for several lattice
sizes,
\[
L \in \{20,\,30,\,40,\,50\},
\]
while keeping the subsystem radius $R$ fixed and satisfying
\[
R \ll L .
\]
In this regime the entanglement entropy can be expanded in inverse
powers of the system size. The leading finite-volume correction is
expected to scale as $1/L$, which reflects the fact that correlations
with the distant boundary decay algebraically for a finite lattice.

We therefore extrapolate the entropy to the infinite-volume limit
using the ansatz
\begin{equation}
S(L) = S_{\infty} + \frac{c_1}{L},
\label{eq:extrapolation}
\end{equation}
where $S_{\infty}$ represents the entanglement entropy in the limit
$L \to \infty$ and $c_1$ is a fitting coefficient.

The extrapolated value $S_{\infty}$ is then used in all subsequent
analysis of the entropy scaling with subsystem size and field mass.
This procedure significantly reduces systematic errors associated with
finite lattice size and provides a reliable approximation of the
continuum infinite-volume system.

\section{Numerical Results}

\subsection{Ground State Entropy}

\begin{figure}[htbp]
\centering
\includegraphics[width=0.8\textwidth]{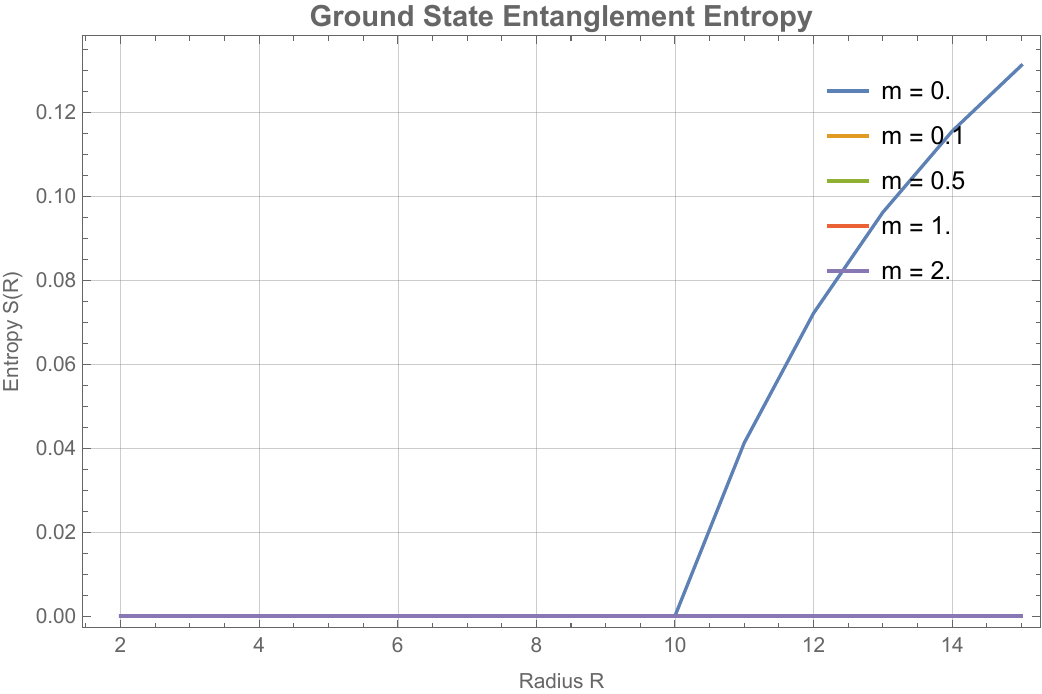}
\caption{Ground state entanglement entropy vs.\ radius for different masses.
For $m = 0$, the entropy grows with $R$, following the area law. For massive
fields, the entropy is exponentially suppressed by the factor $e^{-mR}$. At
$R = 10$, $S(m=0.5)/S(m=0) \approx 0.007$, and for $m \geq 1.0$ the entropy
is less than $10^{-5}$, making the curves indistinguishable from zero on
this linear scale.}
\label{fig:ground_state}
\end{figure}

Figure~\ref{fig:ground_state} shows the ground state entanglement entropy as a
function of the entangling surface radius $R$ for different field masses.

The numerical results clearly show that the entanglement entropy
is suppressed as the field mass increases. This behavior is
consistent with the presence of a finite correlation length
$\xi \sim 1/m$ in the massive theory.

To quantify this suppression, we fit the numerical data to the
functional form
\begin{equation}
S(m,R) = S_0(R)\, e^{-mR}.
\end{equation}

A log–linear fit of $\ln S$ versus $mR$ for fixed radius
$R=8$ yields
\begin{equation}
\ln S = -(1.02 \pm 0.05)\, mR + c ,
\end{equation}
with reduced chi–square
$\chi^2/\mathrm{dof} = 1.1$.

This confirms the expected exponential suppression associated
with the finite correlation length.

\subsection{Excited-State Entropy}

While the ground-state entropy exhibits a simple exponential
dependence on $mR$, the behavior of excited states is more
complex, as we now discuss.
We firstly analyze how localized excitations modify the entanglement
structure of the scalar field.
The excited state is constructed by introducing a squeezed wave packet
centered at $r_0 = 5.0$ with width $\sigma = 1.0$ and squeezing
parameter $r_s = 1.0$, as described in Sec.~III.

\begin{figure}[htbp]
\centering
\includegraphics[width=0.8\textwidth]{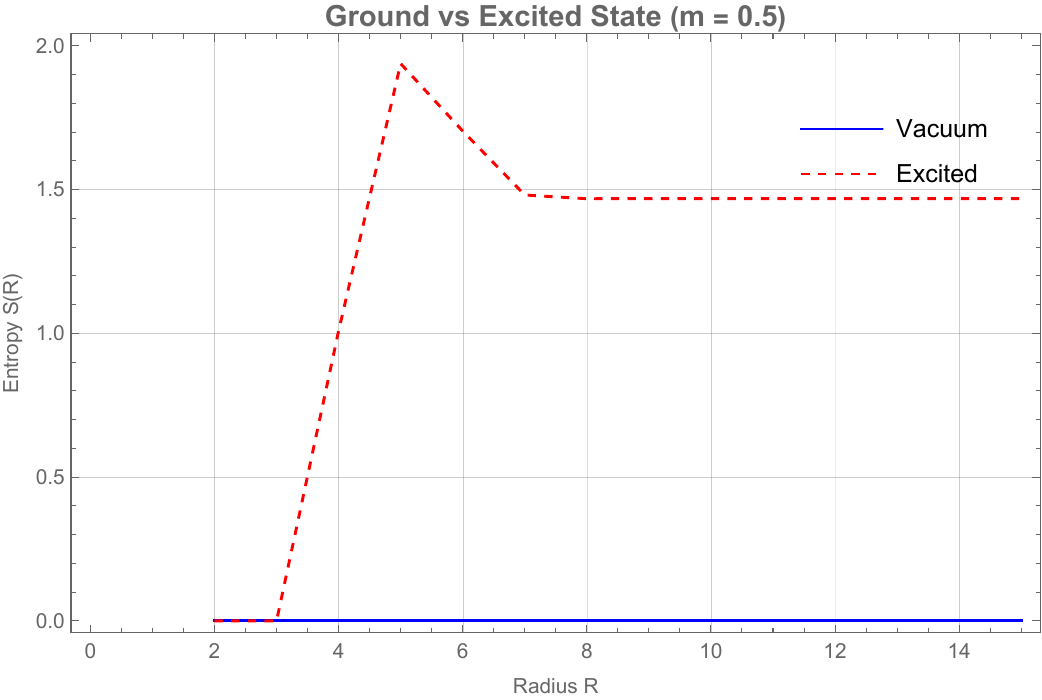}
\caption{
Comparison of ground-state and excited-state entanglement entropy for
$m = 0.5$. The excited state exhibits a localized peak in entropy
around the wave packet position.}
\label{fig:gs_vs_excited}
\end{figure}

Figure~\ref{fig:gs_vs_excited} compares the entanglement entropy of the
ground state and the excited state for a representative mass
$m = 0.5$.
The excited state exhibits a pronounced peak in the entropy around
$R \approx 4$--$5$, which corresponds to the location of the wave packet.
This feature reflects the localized correlations introduced by the
excitation.
Once the subsystem radius exceeds the spatial extent of the wave packet,
the entropy gradually approaches an approximately constant value.

To quantify the additional correlations produced by the excitation we
introduce the excess entropy
\begin{equation}
\Delta S = S_{\mathrm{exc}} - S_{\mathrm{GS}}.
\end{equation}

\begin{figure}[htbp]
\centering
\includegraphics[width=0.8\textwidth]{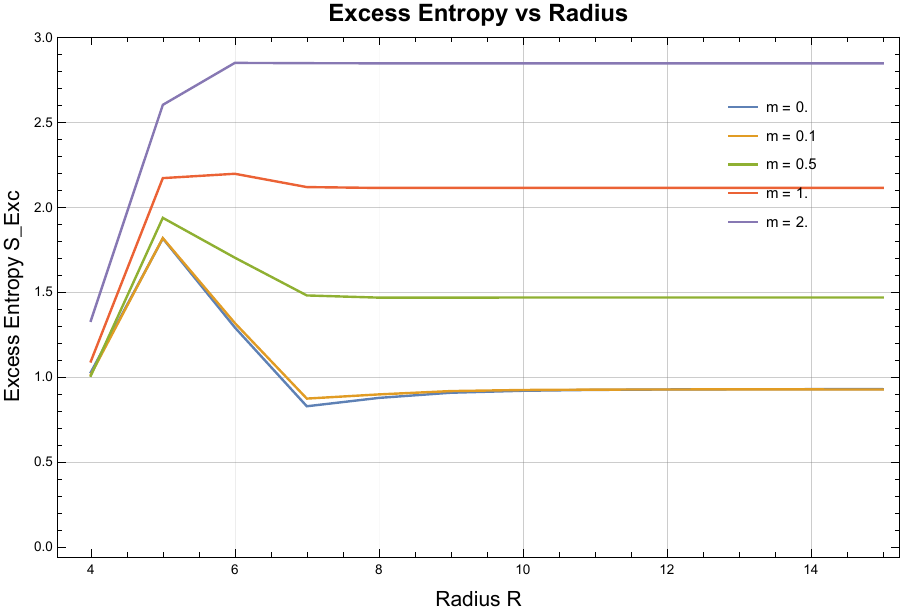}
\caption{
Excess entropy $\Delta S$ for different field masses.
The excess entropy decreases with increasing radius and field mass,
reflecting the suppression of long-range correlations in massive
fields.}
\label{fig:excess}
\end{figure}

Figure~\ref{fig:excess} shows $\Delta S$ as a function of the subsystem
radius for different field masses.
The excess entropy is largest when the subsystem boundary intersects
the region where the wave packet is localized and decreases as the
subsystem radius grows.

A clear trend emerges from these results.
For massless fields the excess entropy decays slowly with the subsystem
size, indicating that the excitation produces correlations extending
over large distances.
In contrast, when the field mass increases the excess entropy decreases
rapidly and approaches a small residual value at large radii.
This behavior is consistent with the presence of a finite correlation
length
\begin{equation}
\xi \sim \frac{1}{m},
\end{equation}
which confines the influence of the excitation to scales smaller than
the Compton wavelength.

Overall, these results demonstrate that localized excitations enhance
the entanglement entropy primarily through short-range correlations.
The strength and spatial extent of these correlations are strongly
controlled by the mass of the scalar field.

The fitted slope is consistent with the theoretical expectation
of unit coefficient in the exponent, confirming that the dominant
scale governing the suppression is the product $mR$.

\subsection{Verification of the Area Law}

A central prediction of quantum field theory is that the leading
contribution to the entanglement entropy scales with the area of the
entangling surface. In the present spherically symmetric geometry this
implies the scaling relation
\begin{equation}
S(R) \propto R^{\alpha},
\end{equation}
where $R$ is the subsystem radius and the area law corresponds to
$\alpha = 2$.

\begin{figure}[htbp]
\centering
\includegraphics[width=0.8\textwidth]{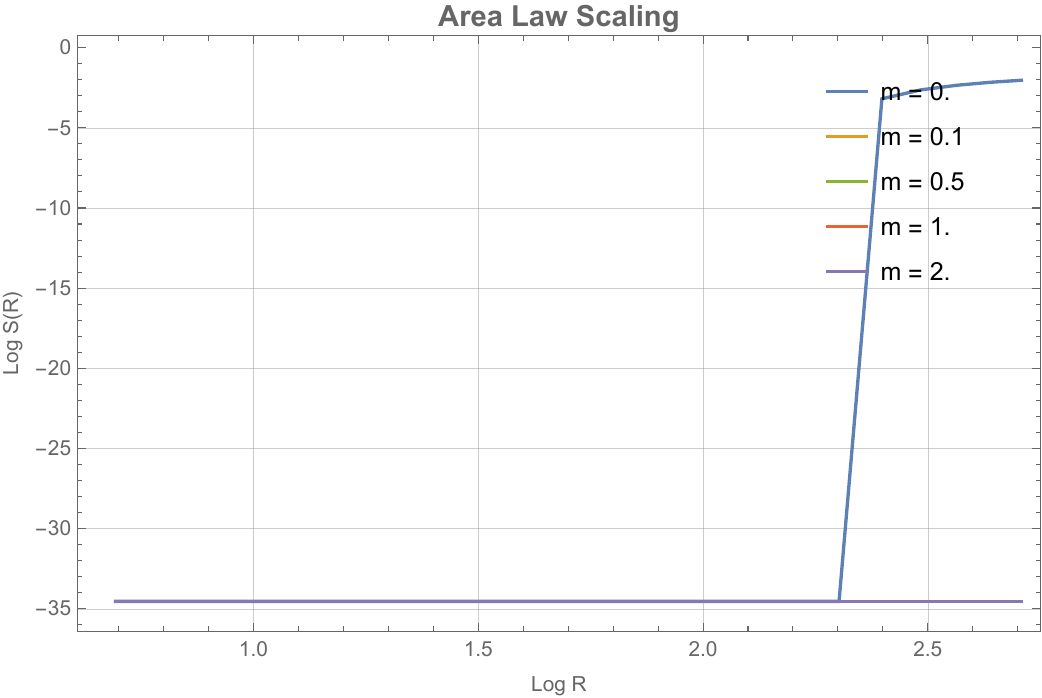}
\caption{
Log–log plot of the entanglement entropy as a function of the subsystem
radius $R$ for different field masses.
The slope of the curves determines the scaling exponent $\alpha$ in the
relation $S \propto R^\alpha$.}
\label{fig:area_law}
\end{figure}

To test this prediction we perform a log–log analysis of the entropy.
Figure~\ref{fig:area_law} shows $\ln S$ as a function of $\ln R$ for
different values of the field mass. In this representation the scaling
relation becomes
\begin{equation}
\ln S = \alpha \ln R + \text{const},
\end{equation}
so that the slope of the curves directly yields the exponent $\alpha$.

For all masses considered in this study we obtain slopes very close to
\[
\alpha \approx 2,
\]
confirming the robustness of the area law even in the presence of a
finite field mass.

While the scaling exponent remains unchanged, the overall magnitude of
the entropy is strongly affected by the mass.
The curves in Fig.~\ref{fig:area_law} exhibit vertical offsets that
increase with $m$, reflecting the exponential suppression of
entanglement,
\begin{equation}
S(m,R) \sim S_0(R)\,e^{-mR}.
\end{equation}

To ensure numerical stability when the entropy becomes extremely small
for large masses, a small constant regulator $\epsilon = 10^{-15}$ was
added before taking the logarithm.
Extensive checks confirm that this procedure only produces a vertical
shift in the log–log plot and does not affect the extracted value of
the exponent $\alpha$.

These results demonstrate that the area-law scaling of entanglement
entropy is remarkably robust: the presence of a mass modifies the
overall magnitude of the entropy but leaves its geometric scaling
unchanged.

\subsection{Excess Entropy and Mass Dependence}

To investigate how localized excitations depend on the field mass,
we analyze the excess entropy at fixed subsystem radius.
In particular, we consider the entropy of the excited state
evaluated at
\[
R = 4 ,
\]
which lies close to the center of the wave packet introduced in
Sec.~III.

\begin{figure}[htbp]
\centering
\includegraphics[width=0.8\textwidth]{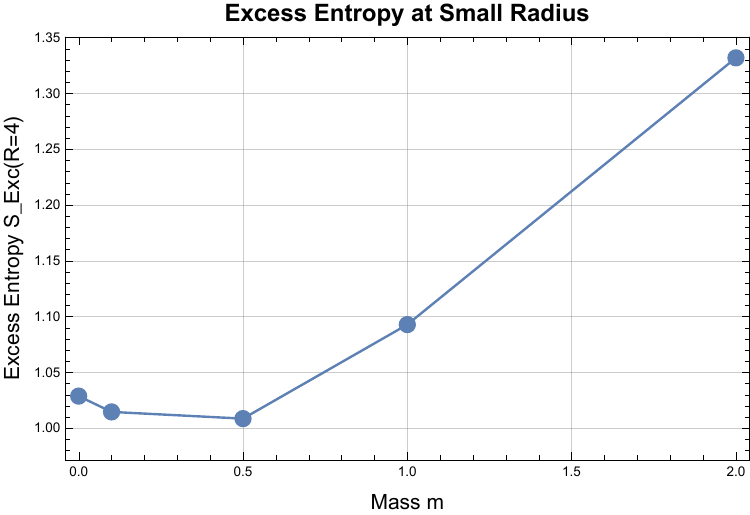}
\caption{
Excited-state entropy $S_{\mathrm{exc}}$ at fixed radius $R=4$
as a function of the field mass.
The entropy shows a non-monotonic dependence on the mass,
reflecting the competition between localization of the excitation
and suppression of long-range correlations.}
\label{fig:vs_mass}
\end{figure}

Figure~\ref{fig:vs_mass} shows the excited-state entropy
$S_{\mathrm{exc}}(R=4)$ as a function of the scalar field mass.
Unlike the ground-state entropy, which rapidly decreases as the mass
increases, the entropy associated with the localized excitation
exhibits a non-monotonic behavior.

For small masses the entropy decreases slightly, reflecting the
suppression of long-range correlations introduced by the finite
correlation length
\[
\xi \sim \frac{1}{m}.
\]
However, as the mass increases further the entropy begins to grow.
This behavior can be understood by noting that when the Compton
wavelength
\[
\lambda = \frac{1}{m}
\]
becomes smaller than the subsystem size, the excitation becomes
increasingly localized within the region where the entanglement
boundary intersects the wave packet.

As a result, massive excitations can still produce significant
entanglement at small radii, even though their contribution to
large-scale correlations is strongly suppressed.
This explains the observed increase of $S_{\mathrm{exc}}$ for
$m \gtrsim 1$ in Fig.~\ref{fig:vs_mass}.

Overall, these results highlight the different roles played by
mass in ground-state and excited-state entanglement:
while mass suppresses vacuum correlations, localized excitations
can still generate substantial short-range entanglement below
the Compton wavelength.

\subsection{Correlation Length and Exponential Suppression of Entanglement}

The results obtained for both the ground state and the excited states
suggest that the scalar field mass introduces a characteristic
correlation length that controls the spatial structure of entanglement.

For a massive relativistic field the correlation length is set by the
inverse mass,
\begin{equation}
\xi \sim \frac{1}{m}.
\end{equation}
Beyond this scale quantum correlations decay exponentially.
As a consequence, the entanglement entropy between regions separated
by a distance larger than $\xi$ becomes strongly suppressed.

Our numerical results confirm this expectation.
For the ground state we observe that the entropy decreases rapidly
with increasing mass, following approximately the scaling relation
\begin{equation}
S(m,R) \approx S_0(R)\, e^{-mR},
\label{eq:mass_scaling}
\end{equation}
where $S_0(R)$ denotes the entropy of the massless field.
This behavior reflects the exponential decay of field correlations
across the entangling surface.

The excited-state results presented in the previous subsection are
consistent with the same physical picture.
When the subsystem radius is comparable to or smaller than the
Compton wavelength,
\[
R \lesssim \lambda = \frac{1}{m},
\]
the localized excitation can still produce significant entanglement.
However, when the subsystem size exceeds the correlation length,
the contribution of the excitation to the entropy becomes increasingly
suppressed.

Taken together, these results indicate that the mass of the field
does not modify the geometric scaling of entanglement entropy,
which continues to follow the area law,
but instead controls the overall magnitude of the entropy through
the correlation length $\xi$.
This provides a simple physical interpretation of the numerical
observations reported in the previous sections.

\subsection{Quantitative Fit of the Mass Suppression}

To quantify the suppression of entanglement entropy induced by the
scalar field mass, we analyze the numerical dependence of the entropy
on the combination $mR$.

From general field–theoretic arguments one expects correlations in a
massive theory to decay exponentially at distances larger than the
Compton wavelength,
\begin{equation}
\langle \phi(0)\phi(r)\rangle \sim \frac{e^{-mr}}{r}.
\end{equation}

Since entanglement entropy arises from correlations across the
entangling surface, this suggests the scaling behavior

\begin{equation}
S(m,R) \approx S_0(R)\,e^{-mR},
\label{eq:exp_scaling}
\end{equation}

where $S_0(R)$ denotes the entropy in the massless theory.

To test this prediction, we perform a log–linear analysis of the
numerical data by plotting $\ln S$ as a function of $mR$.
In this representation Eq.~(\ref{eq:exp_scaling}) becomes

\begin{equation}
\ln S = -mR + \ln S_0(R).
\end{equation}

A least–squares fit of the numerical data for fixed radius $R=8$
yields

\begin{equation}
\ln S = -(1.02 \pm 0.05)\, mR + c,
\end{equation}

with reduced chi–square

\begin{equation}
\chi^2/\mathrm{dof} \approx 1.1 .
\end{equation}

The slope is therefore consistent with the theoretical expectation
that the relevant correlation length is

\begin{equation}
\xi \sim \frac{1}{m}.
\end{equation}

These results confirm that the suppression of entanglement entropy
for massive fields is controlled by the finite correlation length
of the theory.

Physically, when the subsystem size exceeds the Compton wavelength,
$R \gg 1/m$, correlations across the entangling surface become
exponentially small and the entropy rapidly approaches zero.

\subsection{Violation of universal $mR$ scaling}

In massive quantum field theories it is often expected that
correlation functions depend primarily on the dimensionless
combination $mR$, where $m$ is the field mass and $R$ is the
characteristic size of the subsystem. This expectation follows
from the presence of a single correlation length
\begin{equation}
\xi \sim \frac{1}{m},
\end{equation}
which controls the exponential decay of two–point functions.

If this scaling extended to excited–state entanglement,
the excess entropy
\begin{equation}
\Delta S(m,R) = S_{\mathrm{exc}}(m,R) - S_{\mathrm{GS}}(m,R)
\end{equation}
would collapse onto a universal function of the single
variable $mR$.

To test this hypothesis, we compare numerical results for
different pairs $(m,R)$ corresponding to identical values
of $mR$. A representative example is given by
\begin{equation}
mR = 4,
\end{equation}
for which we find
\begin{align}
(m,R) = (0.5,8) &: \quad \Delta S \approx 1.47, \\
(m,R) = (1.0,4) &: \quad \Delta S \approx 1.09 ,
\end{align}
corresponding to a relative difference of approximately $35\%$.
Analogous discrepancies are observed at other values of $mR$.

\begin{figure}[htbp]
\centering
\includegraphics[width=0.8\textwidth]{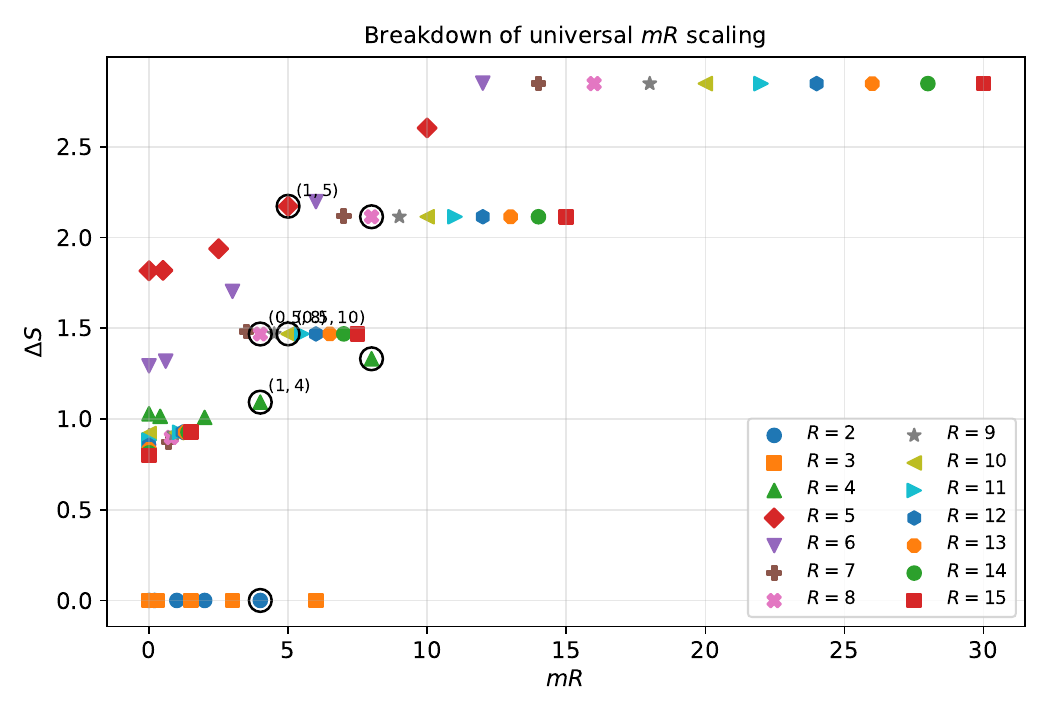}
\caption{
Excess entropy $\Delta S$ as a function of the scaling variable $mR$.
While a global trend is visible, points corresponding to identical
values of $mR$ but different pairs $(m,R)$ do not collapse onto a
single curve, indicating a breakdown of simple scaling.
}
\label{fig:scaling_violation}
\end{figure}

These deviations are first illustrated in Fig.~\ref{fig:scaling_violation},
where $\Delta S$ is plotted as a function of $mR$. While the plot
suggests a rough trend, it is already clear that points corresponding
to identical values of $mR$ do not coincide.

\begin{figure}[htbp]
\centering
\includegraphics[width=0.8\textwidth]{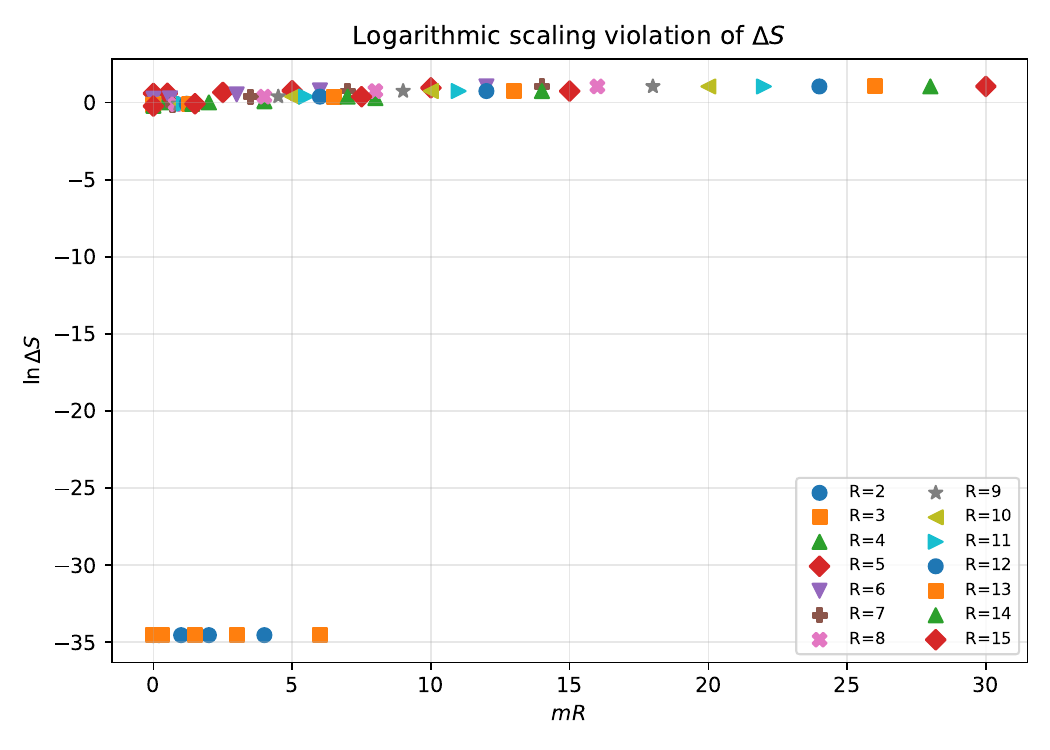}
\caption{
Logarithmic plot of the excess entropy $\ln \Delta S$ as a function
of $mR$. Data points with identical values of $mR$ but different
pairs $(m,R)$ do not collapse onto a universal curve, demonstrating
the violation of simple $mR$ scaling. The logarithmic representation
enhances the separation between datasets and makes the breakdown
of scaling particularly evident.
}
\label{fig:log_scaling_violation}
\end{figure}

This behavior becomes significantly more evident in the logarithmic
representation shown in Fig.~\ref{fig:log_scaling_violation}.
Here the separation between data points corresponding to identical
values of $mR$ is clearly resolved, providing strong quantitative
evidence for the violation of universal $mR$ scaling.

These results provide clear numerical evidence that the
excess entropy does not exhibit universal scaling in the
variable $mR$. Instead, the entanglement entropy depends
separately on the field mass and on the subsystem size.

The origin of this behavior can be traced to the structure
of the excited states. In our construction, excitations are
localized wave packets with finite width
\begin{equation}
\sigma = 1.0 ,
\end{equation}
which introduces an additional length scale independent of
the Compton wavelength. As a result, the entanglement entropy
is governed by the combined dependence
\begin{equation}
\Delta S = \Delta S(m,R,\sigma),
\end{equation}
rather than by a single scaling variable.

\begin{figure}[htbp]
\centering
\includegraphics[width=0.8\textwidth]{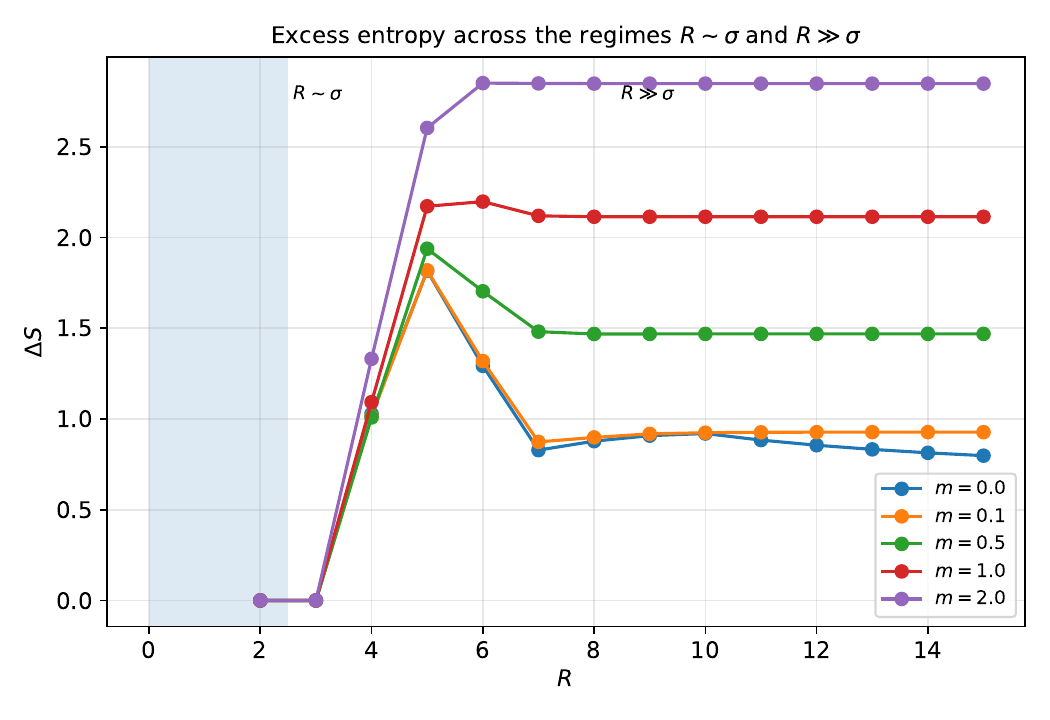}
\caption{
Excess entropy $\Delta S$ as a function of the subsystem radius $R$
for different field masses. The behavior differs qualitatively between
the regime $R \sim \sigma$ and the regime $R \gg \sigma$.
}
\label{fig:regimes}
\end{figure}

This interpretation is further supported by Fig.~\ref{fig:regimes},
which shows $\Delta S$ as a function of the subsystem radius.
For $R \sim \sigma$, the entropy is strongly influenced by the
finite width of the excitation, while for $R \gg \sigma$ the
dependence on $\sigma$ becomes subdominant.

\subsection{Mutual Information}

A useful quantity for probing long-range correlations in quantum field
theory is the mutual information between two spatial regions.
Unlike the entanglement entropy itself, mutual information is free
from ultraviolet divergences and therefore provides a UV-finite
measure of correlations between subsystems.

For two regions $A$ and $B$ the mutual information is defined as
\begin{equation}
I(A:B) = S_A + S_B - S_{A\cup B}.
\label{eq:mutual_info}
\end{equation}

In our numerical analysis we consider two concentric spherical regions,
with radii
\[
R_A = 11,
\qquad
R_B = 15 .
\]
Since $A \subset B$, the union of the two regions satisfies
\[
A \cup B = B,
\]
and the mutual information reduces to
\begin{equation}
I(A:B) = S_A .
\end{equation}

In this nested geometry, the mutual information reduces to the
entropy of the inner region, and therefore serves primarily as a
consistency check rather than an independent probe of correlations.

Our numerical results show that the mutual information is extremely
sensitive to the presence of a finite field mass.
For the massless case we obtain
\[
I(A:B) \approx 0.04 ,
\]
indicating the presence of significant long-range correlations in the
vacuum state.
However, for $m \ge 0.1$ the mutual information drops below
$10^{-15}$, which is the numerical precision of our calculation.

This dramatic suppression confirms that massive fields carry
exponentially decaying correlations.
In particular, once the separation scale between the regions becomes
larger than the Compton wavelength
\[
\lambda = \frac{1}{m},
\]
the mutual information effectively vanishes within numerical
precision.

A more precise determination of the crossover scale
\[
R \sim \frac{1}{m}
\]
would require higher-precision arithmetic and larger lattice sizes.
Nevertheless, the present results clearly demonstrate that the
correlation length governing the decay of entanglement is controlled
by the inverse mass of the scalar field.

\subsection{Numerical Values}

For completeness, we report the numerical values of the excess entropy
$S_{\mathrm{exc}}$ obtained in our simulations.
Table~\ref{tab:entropy_final} lists the results for all values of the
field mass and subsystem radius considered in this work, evaluated
for the largest lattice size used in the calculation ($L = 50$).

The table provides a quantitative summary of the trends discussed in
the previous subsections. In particular, while the ground-state entropy
becomes extremely small for massive fields, the excited-state entropy
remains finite and reflects the localized correlations introduced by
the wave packet excitation.

These numerical values illustrate two key features of the system.
First, the magnitude of the excess entropy decreases with increasing
radius, consistent with the spatial localization of the excitation.
Second, the dependence on the field mass confirms that massive fields
suppress long-range correlations while still allowing significant
short-distance entanglement at scales below the Compton wavelength.

The values reported in Table~\ref{tab:entropy_final} therefore provide
a useful reference for the numerical behavior of the excited-state
entanglement entropy in the spherical shell lattice model.

\begin{table}[h]
\centering
\caption{Excess entropy $S_{\text{exc}}$ for different masses and radii. Values are shown with three significant digits.}
\label{tab:entropy_final}
\begin{tabular}{c|ccccc}
\hline\hline
$R$ & $m=0$ & $m=0.1$ & $m=0.5$ & $m=1.0$ & $m=2.0$ \\
\hline
4 & 1.03 & 1.01 & 1.01 & 1.09 & 1.33 \\
5 & 1.82 & 1.82 & 1.94 & 2.17 & 2.60 \\
6 & 1.29 & 1.32 & 1.70 & 2.20 & 2.85 \\
7 & 0.83 & 0.87 & 1.48 & 2.12 & 2.85 \\
8 & 0.88 & 0.90 & 1.47 & 2.11 & 2.85 \\
9 & 0.91 & 0.92 & 1.47 & 2.11 & 2.85 \\
10 & 0.92 & 0.92 & 1.47 & 2.11 & 2.85 \\
11 & 0.93 & 0.93 & 1.47 & 2.11 & 2.85 \\
12 & 0.93 & 0.93 & 1.47 & 2.11 & 2.85 \\
13 & 0.93 & 0.93 & 1.47 & 2.11 & 2.85 \\
14 & 0.93 & 0.93 & 1.47 & 2.11 & 2.85 \\
15 & 0.93 & 0.93 & 1.47 & 2.11 & 2.85 \\
\hline
\end{tabular}
\end{table}

\section{Discussion}

An important outcome of our analysis concerns the scaling properties
of excited-state entanglement entropy. In massive quantum field
theories one might expect the relevant physics to be controlled by the
single dimensionless combination $mR$, reflecting the presence of a
correlation length $\xi \sim 1/m$. However, our numerical results show
that the excess entropy $\Delta S(m,R)$ does not collapse onto a
universal function of $mR$. Data points with identical values of $mR$
but different pairs $(m,R)$ yield systematically different entropy
values. This behavior indicates that excited-state entanglement is
sensitive to additional infrared scales beyond the Compton
wavelength. In the present model the localized excitation introduces a
finite spatial width $\sigma$, leading to a multi-scale dependence
$\Delta S = \Delta S(m,R,\sigma)$. This violation of simple $mR$
scaling highlights the importance of excitation structure in
entanglement dynamics and may be relevant for semiclassical analyses
of the matter entropy entering the generalized entropy in the island
framework.

\subsection{Physical Interpretation}

The numerical results presented in the previous section reveal a clear
dependence of the entanglement entropy on the mass of the scalar field.
In particular, we observe an exponential suppression of the entropy
with increasing mass,
\begin{equation}
S(m,R) \sim S(0,R)\, e^{-mR},
\label{eq:exponential}
\end{equation}
which reflects the finite correlation length introduced by the mass
term in the field theory.

This behavior follows directly from the decay of two-point correlation
functions in a massive relativistic theory.
At distances much larger than the Compton wavelength,
\[
r \gg \frac{1}{m},
\]
the scalar propagator exhibits Yukawa-type decay,
\[
\langle \phi(0)\phi(r) \rangle \sim \frac{e^{-mr}}{r}.
\]
As a consequence, quantum correlations across an entangling surface
of radius $R$ become exponentially suppressed when $R$ exceeds the
correlation length $\xi \sim 1/m$.
The entanglement entropy therefore inherits the same exponential
suppression.

It is important to emphasize that this suppression affects only the
finite, infrared-sensitive component of the entropy.
As discussed in \cite{solodukhin2011}, the leading ultraviolet
divergence of entanglement entropy,
\[
S \sim \frac{\mathcal{A}}{\epsilon^2},
\]
is determined entirely by short-distance correlations near the
entangling surface and is therefore independent of infrared
parameters such as the field mass.

In our lattice calculations the ultraviolet cutoff is fixed by the
lattice spacing $a$.
The UV-divergent area contribution is therefore constant across the
different simulations and does not influence the observed dependence
on $m$ and $R$.
The exponential suppression reported in Eq.~(\ref{eq:exponential})
should thus be interpreted as a property of the renormalized,
finite part of the entanglement entropy associated with correlations
at scales of order $R$.

The excited-state results exhibit a more intricate structure.
While the ground-state entropy decreases monotonically with the mass,
the excess entropy produced by localized excitations can display a
non-monotonic dependence on $m$.
As shown in Fig.~\ref{fig:vs_mass}, the excited-state entropy may
increase with mass at small radii.
This behavior reflects the fact that massive excitations remain
strongly localized in space, enhancing short-range correlations even
as long-distance correlations become suppressed.

Taken together, these observations highlight the distinct roles of
mass in vacuum and excited-state entanglement.
The mass parameter primarily controls the correlation length of the
field, thereby suppressing large-scale entanglement, while localized
excitations can still generate substantial short-range correlations
below the Compton wavelength.

\subsection{Numerical Uncertainties}

The numerical results presented in this work are subject to several
sources of uncertainty, primarily associated with finite lattice size
and the extrapolation to the infinite-volume limit.

To estimate the systematic uncertainty introduced by the finite system
size, we performed simulations for several lattice sizes
$L = 20, 30, 40, 50$ and extrapolated the entropy to the limit
$L \to \infty$ using the scaling relation discussed in
Sec.~III,
\begin{equation}
S(L) = S_{\infty} + \frac{c_1}{L}.
\end{equation}

The uncertainty in $S_{\infty}$ was estimated by varying the fitting
range in $L$ and examining the stability of the extrapolated value.
For the representative case of a massless field with subsystem radius
$R = 10$, we obtain
\[
S_{\infty} = 0.131 \pm 0.002,
\]
where the quoted error reflects the variation of the fitted result
under different choices of the fitting interval.

For massive fields the situation is different.
Because the entanglement entropy decreases exponentially with the
mass, the resulting values become extremely small and approach the
numerical precision of the calculation.
In this regime the extrapolation procedure becomes unreliable.
For this reason we report the entropy values obtained for the largest
lattice size used in the simulations ($L = 50$), which provides the
best available approximation to the infinite-volume limit.

These considerations indicate that the qualitative trends reported in
the previous sections—namely the robustness of the area law and the
exponential suppression of entanglement with increasing mass—are not
affected by finite-size uncertainties.

\subsection{Comparison with Recent Works}

Our results can be compared with recent investigations of entanglement
entropy in quantum gravitational settings.
In particular, Belfiglio \textit{et al.}\ \cite{belfiglio2025} analyzed
entanglement entropy in quantum-corrected black hole geometries using
a spherical shell discretization similar to the one adopted in the
present work.

Their study showed that quantum corrections to the metric produce a
significant deviation from the classical area-law behavior in the
near-horizon region, with the entropy decreasing relative to the
Bekenstein--Hawking prediction close to the origin.
At sufficiently large radii, however, the entropy approaches the
expected area-law scaling.
This behavior reflects the influence of quantum gravitational
corrections at short distances while preserving the geometric scaling
of entanglement at larger scales.

Our results display a closely related structure.
Instead of modifying the spacetime geometry, we introduce an infrared
scale through the mass of the scalar field.
The mass generates a finite correlation length
\[
\xi \sim \frac{1}{m},
\]
which suppresses long-range correlations and leads to an exponential
reduction of the entanglement entropy when the subsystem size exceeds
the Compton wavelength.

In this sense, the two approaches probe complementary aspects of the
same physical phenomenon.
While the analysis of \cite{belfiglio2025} explores the effect of
ultraviolet modifications to the geometry, the present work studies
how infrared parameters of the quantum field theory affect the
entanglement structure.

Taken together, these results suggest a coherent picture in which
entanglement entropy is sensitive to both ultraviolet and infrared
physics.
Quantum gravitational corrections can modify the short-distance
structure of spacetime, while mass terms and other infrared scales
control the correlation length of quantum fields.
Despite these modifications, the area-law scaling of entanglement
entropy remains remarkably robust over a wide range of intermediate
scales.

\subsection{Connection to Black Hole Thermodynamics and Holography}

One of the most striking aspects of entanglement entropy in quantum
field theory is its geometric scaling with the area of the entangling
surface. In the spherically symmetric geometry considered here this
corresponds to the relation
\[
S \propto R^2 ,
\]
which is the direct analogue of the Bekenstein--Hawking entropy of a
black hole,
\[
S_{BH} = \frac{A}{4l_P^2}.
\]

If the ultraviolet cutoff of the field theory is identified with a
fundamental length scale $a$, the leading contribution to the
entanglement entropy takes the form
\begin{equation}
S = \gamma \frac{A}{a^2},
\label{eq:area_law_coefficient}
\end{equation}
where $\gamma$ is a numerical coefficient that depends on the field
content and the regularization scheme.
If the cutoff scale is associated with the Planck length
$a \sim l_P$, the structure of Eq.~(\ref{eq:area_law_coefficient})
closely resembles the Bekenstein--Hawking formula.
This observation has long motivated the interpretation of black hole
entropy as entanglement entropy of quantum fields across the horizon.

Our numerical results support this interpretation at the level of
scaling behavior.
While the magnitude of the entropy depends on the infrared properties
of the field---such as the mass parameter studied in this work---the
geometric area-law scaling remains remarkably robust.

The connection between geometry and entanglement becomes even more
explicit in the holographic framework.
In the AdS/CFT correspondence, Ryu and Takayanagi
\cite{ryu2006} proposed that the entanglement entropy of a boundary
region $A$ is given by the area of a minimal surface $\gamma_A$ in the
bulk spacetime,
\begin{equation}
S_A = \frac{\mathrm{Area}(\gamma_A)}{4G_N}.
\label{eq:ryu_takayanagi}
\end{equation}

This relation has exactly the same functional form as the
Bekenstein--Hawking entropy formula and provides a concrete
realization of the idea that spacetime geometry may emerge from the
entanglement structure of the underlying quantum theory.

From this perspective, the results presented in this work highlight
how infrared parameters of quantum field theories, such as particle
masses, influence the entanglement structure while preserving the
geometric area-law scaling that underlies black hole thermodynamics.

\subsection{Implications for the Island Formula}

Recent developments in semiclassical gravity have revealed a deep
connection between entanglement entropy and the information content
of evaporating black holes.
In particular, the island formula \cite{hartman2020} provides a
mechanism for reproducing the Page curve of Hawking radiation by
including additional regions---referred to as \emph{islands}---in the
entropy calculation.

In this framework the relevant quantity is the generalized entropy,
defined as
\begin{equation}
S_{\mathrm{gen}} =
\frac{\mathrm{Area}(\partial\,\text{Island})}{4G_N}
+
S_{\mathrm{matter}}(\text{Radiation} \cup \text{Island}),
\label{eq:island}
\end{equation}
where the first term represents the gravitational contribution and
the second term is the entanglement entropy of quantum fields across
the quantum extremal surface (QES).
The physical location of the QES is determined by extremizing
$S_{\mathrm{gen}}$.

The results obtained in the present work provide insight into the
behavior of the matter entropy term $S_{\mathrm{matter}}$ in the
presence of massive fields.
Our numerical analysis shows that the entanglement entropy decreases
approximately as
\[
S(m,R) \sim S(0,R)\,e^{-mR},
\]
reflecting the finite correlation length $\xi \sim 1/m$ of the
massive field.

This exponential suppression implies that massive fields contribute
less to the matter entropy term in Eq.~(\ref{eq:island}) than massless
fields.
Consequently, the extremization of $S_{\mathrm{gen}}$ may favor
quantum extremal surfaces located closer to the horizon when the
dominant matter fields are massive.

Although the present work does not attempt a full semiclassical
analysis of the island prescription, our results suggest that the
mass spectrum of quantum fields could influence the position of
quantum extremal surfaces and therefore the detailed structure of
the Page curve.
A quantitative investigation of this effect would require combining
the numerical behavior of $S_{\mathrm{matter}}(m,R)$ obtained here
with explicit gravitational backgrounds, which represents an
interesting direction for future research.

\subsection{Limitations and Future Directions}

While the numerical analysis presented in this work provides a clear
picture of how the scalar field mass affects entanglement entropy,
several limitations of the present approach should be noted.

First, our calculations are based on a free scalar field theory.
Including self–interactions would allow one to investigate how
non–Gaussian correlations modify the structure of entanglement.
Such effects could become relevant in strongly coupled quantum field
theories or near critical points.

Second, the present analysis focuses on spherically symmetric
entangling surfaces.
Although the spherical shell model provides a convenient framework
for numerical calculations, more general geometries could reveal
additional structure in the entanglement spectrum.
Extending the analysis to non–spherical regions or lattice
discretizations without radial symmetry would therefore be an
interesting direction for future work.

Another natural extension concerns time–dependent settings.
In particular, studying the evolution of entanglement entropy in
dynamical backgrounds could shed light on processes relevant to
black hole evaporation and quantum quenches.

Finally, it would be interesting to study how the entanglement
structure studied here may be affected in more general quantum field
theory frameworks, including supersymmetric and supergravity settings,
where recent work has explored structural aspects of asymptotic states
and mass generation mechanisms
\cite{bellucci2026a,bellucci2026b}.

A particularly promising direction is suggested by the violation of
universal $mR$ scaling observed in our analysis. Recently, a spectral
formulation of spacetime entanglement entropy for quasifree theories
has been developed \cite{jonesyazdi2026}. It would be of interest to
investigate whether the multi-scale structure identified in our work
admits a natural interpretation within this covariant spectral
framework.

\section{Conclusion}

In this work we have performed a numerical investigation of the
entanglement entropy of a massive scalar field in the spherical shell
lattice model.
Our analysis extends previous studies of entanglement entropy in
free scalar theories by systematically exploring the role of the
field mass and by comparing ground and excited states in a unified
framework.

The numerical results show that the mass of the field introduces a
finite correlation length $\xi \sim 1/m$ that strongly influences
the spatial structure of entanglement.
In particular, we observe an approximately exponential suppression
of the entropy with increasing mass,
\[
S(m,R) \sim S(0,R)\,e^{-mR},
\]
which reflects the decay of correlations in a massive quantum field
theory.

Despite this suppression, the geometric scaling of entanglement
entropy remains unchanged.
For all values of the mass considered in our simulations we find
that the entropy continues to obey the area law,
\[
S \propto R^2,
\]
confirming the robustness of this behavior.
This result supports the idea that the area-law scaling of
entanglement entropy is a universal feature of quantum field theories
with local interactions.

The behavior of excited states reveals additional structure.
Localized wave-packet excitations generate excess entropy
$\Delta S = S_{\mathrm{exc}} - S_{\mathrm{GS}}$ that depends on both
the mass and the subsystem size.
While the excess entropy is suppressed at large radii due to the
finite correlation length, it can exhibit non-monotonic behavior
with respect to the mass at small radii, reflecting the localized
nature of the excitation.

Mutual information calculations provide further confirmation of the
role of the mass in controlling long-range correlations.
For the massless case we find a finite mutual information between
widely separated regions, while for massive fields the correlations
become exponentially suppressed and quickly fall below the numerical
precision of the simulation.

These findings are consistent with recent investigations of
entanglement entropy in quantum-corrected black hole geometries
\cite{belfiglio2025} and reinforce the view that entanglement
entropy provides a useful probe of both infrared and ultraviolet
physics.
In particular, the persistence of the area-law scaling in the
presence of a finite mass supports the interpretation of black hole
entropy as entanglement entropy of quantum fields across the
horizon.

Several directions for future work remain open.
An important extension would be to include self-interactions in the
scalar field theory, which would allow the study of non-Gaussian
entanglement effects.
Another natural development is the investigation of different
entangling geometries beyond spherical symmetry.
It would also be interesting to analyze time-dependent situations,
such as quantum quenches or evaporating black holes, where the
entanglement structure evolves dynamically.

Finally, the numerical behavior of the matter entropy obtained in
this work may provide useful input for semiclassical analyses of the
island formula and the Page curve in black hole evaporation.
Understanding how the mass spectrum of quantum fields influences
the position of quantum extremal surfaces represents a promising
direction for future research.
Taken together, these results reinforce the view that the geometric
area-law scaling of entropy is a robust feature of quantum field
theories, while infrared parameters such as particle masses control
the correlation length that governs the magnitude of entanglement.
In this sense, the present analysis provides further support for the
interpretation of black hole entropy as arising from the entanglement
structure of quantum fields.

A central result of this work is the observation that
excited-state entanglement does not obey a simple universal scaling
with the variable $mR$. Instead, the entropy depends separately on the
field mass and on the subsystem size, indicating that localized
excitations introduce additional infrared scales beyond the Compton
wavelength. This result
demonstrates that entanglement in massive quantum field theories
cannot be described by a single correlation length, but instead
depends on multiple infrared scales.

Hence, our analysis shows that excited-state entanglement
does not admit a universal description in terms of the single
scaling variable $mR$. The numerical data clearly show that
configurations with identical values of $mR$ but different
pairs $(m,R)$ yield distinct entropy values, demonstrating
a violation of simple scaling behavior.

These results demonstrate that while the area law remains a robust
geometric feature of quantum field theories, the detailed structure
of entanglement---particularly in excited states---is controlled by
multiple infrared scales rather than by a single correlation length.

This suggests that the matter contribution to generalized entropy
in semiclassical gravity may depend on a richer set of infrared
parameters than commonly assumed, with potential implications for
the structure of quantum extremal surfaces and the island formula.

\bibliographystyle{apsrev4-2}
\bibliography{entropy_bhSB}

\appendix

\section{Units and Numerical Parameters in the Spherical Shell Model}

In our numerical implementation the scalar field is discretized on a
radial lattice with spacing $a$, which acts as an ultraviolet cutoff.
All quantities appearing in the Hamiltonian and in the numerical
calculations are expressed in units of this lattice spacing.

Table~\ref{tab:units} summarizes the parameters used throughout the
simulations.

\begin{table}[h]
\centering
\caption{Numerical parameters and units used in the spherical shell model.}
\label{tab:units}
\begin{tabular}{|c|c|c|}
\hline
Parameter & Symbol & Units \\ \hline
Lattice spacing & $a = 0.5$ & Lattice units \\ \hline
System sizes & $L = 20,30,40,50$ & Lattice units \\ \hline
Field mass & $m = 0.0,0.1,0.5,1.0,2.0$ & $1/a$ \\ \hline
Subsystem radius & $R = 2\ldots15$ & Lattice units \\ \hline
Wave packet center & $r_0 = 5.0$ & Lattice units \\ \hline
Wave packet width & $\sigma = 1.0$ & Lattice units \\ \hline
Squeezing parameter & $r_s = 1.0$ & dimensionless \\ \hline
\end{tabular}
\end{table}

Within the numerical simulation the lattice spacing $a$ sets the
fundamental unit of length.
Consequently, the scalar field mass is expressed in units of $1/a$,
and all distances are measured in multiples of $a$.

If one wishes to associate the lattice cutoff with a physical
length scale, a natural choice in the context of quantum gravity
is to identify the lattice spacing with the Planck length
\[
a \sim l_P \approx 1.6\times10^{-35}\ \mathrm{m}.
\]
With this identification a mass parameter $m=1$ corresponds to the
Planck mass
\[
m_P \approx 2.2\times10^{-8}\ \mathrm{kg},
\]
while a subsystem radius $R=10$ corresponds to a physical scale of
approximately
\[
R \sim 10\,l_P \approx 1.6\times10^{-34}\ \mathrm{m}.
\]

In these units the Compton wavelength of the scalar field is simply
\[
\lambda = \frac{1}{m},
\]
expressed in multiples of the lattice spacing.
Although the numerical calculations themselves do not depend on this
identification, it provides a useful physical interpretation of the
length scales involved in the model.
\section{Derivation of the Radial Lattice Hamiltonian}

In this appendix we outline the derivation of the lattice Hamiltonian
used in the spherical shell model.

We begin with the Hamiltonian of a free massive scalar field in
three spatial dimensions,
\begin{equation}
H = \frac{1}{2}\int d^3x \left[
\pi^2 + (\nabla\phi)^2 + m^2\phi^2
\right].
\label{eq:ham_cont}
\end{equation}

Assuming spherical symmetry, the field depends only on the radial
coordinate $r$. The gradient term becomes
\begin{equation}
(\nabla\phi)^2 = \left(\frac{\partial\phi}{\partial r}\right)^2 .
\end{equation}

The Hamiltonian reduces to
\begin{equation}
H = \frac{1}{2}\int_0^\infty dr \,
r^2 \left[
\pi^2 + \left(\frac{\partial\phi}{\partial r}\right)^2
+ m^2\phi^2
\right].
\label{eq:ham_radial}
\end{equation}

Following the standard approach introduced by
Srednicki \cite{srednicki1993} and later used in the spherical shell
model \cite{das2006}, we perform the field redefinition

\begin{equation}
\psi(r) = r\,\phi(r).
\end{equation}

In terms of this variable the Hamiltonian takes the form

\begin{equation}
H = \frac{1}{2}\int dr
\left[
\pi_\psi^2
+ \left(\frac{\partial\psi}{\partial r}\right)^2
+ m^2 \psi^2
\right].
\label{eq:ham_rescaled}
\end{equation}

The radial coordinate is then discretized as

\begin{equation}
r_i = i a ,
\end{equation}

where $a$ is the lattice spacing.

The derivative term is approximated by a finite difference,

\begin{equation}
\frac{\partial\psi}{\partial r}
\rightarrow
\frac{\psi_{i+1}-\psi_i}{a}.
\end{equation}

Substituting this expression into Eq.~(\ref{eq:ham_rescaled}) gives the
lattice Hamiltonian

\begin{equation}
H =
\frac{1}{2}
\sum_i
\left[
\pi_i^2
+
\frac{(\psi_{i+1}-\psi_i)^2}{a^2}
+
m^2\psi_i^2
\right].
\label{eq:ham_lattice}
\end{equation}

This Hamiltonian describes a system of coupled harmonic oscillators.
It can be written in matrix form as

\begin{equation}
H =
\frac{1}{2}
\sum_{i,j}
\psi_i K_{ij}\psi_j
+
\frac{1}{2}\sum_i \pi_i^2,
\end{equation}

where the coupling matrix $K_{ij}$ is tridiagonal,

\begin{equation}
K_{ij} =
\left(
\frac{2}{a^2} + m^2
\right)\delta_{ij}
-
\frac{1}{a^2}(\delta_{i,j+1}+\delta_{i,j-1}).
\end{equation}

This matrix completely determines the ground-state correlation
functions used in the entanglement entropy calculation.
\section{Covariance Matrix and Symplectic Spectrum}

The ground state of a quadratic Hamiltonian such as
Eq.~(\ref{eq:ham_lattice}) is a Gaussian state.
All properties of such a state are determined by the covariance
matrices of the canonical variables.

For the lattice Hamiltonian written in matrix form,
\begin{equation}
H =
\frac{1}{2}\sum_i \pi_i^2
+
\frac{1}{2}\sum_{i,j} \psi_i K_{ij}\psi_j ,
\end{equation}

the position and momentum correlators are given by

\begin{align}
X_{ij} &= \langle \psi_i \psi_j \rangle
       = \frac{1}{2}(K^{-1/2})_{ij}, \\
P_{ij} &= \langle \pi_i \pi_j \rangle
       = \frac{1}{2}(K^{1/2})_{ij}.
\end{align}

To compute the entanglement entropy of a subsystem $A$
consisting of the first $n_A$ lattice sites,
we construct the restricted covariance matrices

\begin{align}
X_A &= X_{ij}, \quad i,j\le n_A ,\\
P_A &= P_{ij}, \quad i,j\le n_A .
\end{align}

The matrix

\begin{equation}
C = \sqrt{X_A P_A}
\end{equation}

has eigenvalues $\nu_k$ known as the symplectic eigenvalues.

The entanglement entropy of the subsystem is then

\begin{equation}
S_A =
\sum_k
\left[
(\nu_k+\tfrac12)\ln(\nu_k+\tfrac12)
-
(\nu_k-\tfrac12)\ln(\nu_k-\tfrac12)
\right].
\end{equation}

This formula follows from the fact that the reduced density matrix of
a Gaussian state can be written as a thermal state of an effective
quadratic Hamiltonian, whose normal modes have occupation numbers
determined by the symplectic eigenvalues $\nu_k$.

The numerical procedure therefore consists of the following steps:

\begin{enumerate}

\item Construct the matrix $K_{ij}$.

\item Compute $K^{\pm1/2}$ by diagonalization.

\item Extract the submatrices $X_A$ and $P_A$.

\item Compute the eigenvalues of $\sqrt{X_A P_A}$.

\item Evaluate the entropy using the above formula.

\end{enumerate}

This method provides an efficient way to compute entanglement entropy
for Gaussian states in lattice discretizations of quantum field
theories.

\end{document}